\documentclass[twocolumn,superscriptaddress, floatfix, amsfonts, aps]{revtex4-1} 
\usepackage{amsmath}
\usepackage{bm}
\usepackage{graphicx}
\usepackage{subfigure}
\usepackage{float}
\usepackage{mathrsfs}
\usepackage{footmisc}
\usepackage{multirow}
\usepackage{graphicx}
\usepackage{booktabs}
\usepackage{soul,xcolor}
\usepackage{xcolor, ulem, color, framed}
\usepackage{amssymb}
\usepackage[colorlinks, linkcolor=red,anchorcolor=green,citecolor=blue]{hyperref}

\newcommand{\tred}{\textcolor{black}}

\usepackage[colorlinks, linkcolor=red,anchorcolor=green,citecolor=blue]{hyperref}

\graphicspath{{./figs/}}

\begin{document}

 \title{Exploring Spin Polarization of Heavy Quarks in Magnetic Fields and Hot Medium}

\author{Zhiwei Liu}
\affiliation{Department of Physics, Tianjin University, Tianjin 300354, China}

\author{Yunfan Bai}\email{ab3921@ic.ac.uk}
\affiliation{Department of Physics, Imperial College London, London, SW72AZ, United Kindom}
\author{Shiqi Zheng}\email{shiqi\_zheng@brown.edu}
\affiliation{School of Engineering, Brown University, Providence, RI 02912, USA}

\author{Anping Huang}\email{huanganping@ucas.ac.cn}
\affiliation{School of Material Science and Physics, China University of Mining and Technology, Xuzhou, China}

\author{Baoyi Chen}\email{baoyi.chen@tju.edu.cn}
\affiliation{Department of Physics, Tianjin University, Tianjin 300354, China}

\begin{abstract}
Relativistic heavy-ion collisions give rise to the formation of both deconfined QCD matter and a strong magnetic field. The spin of heavy quarks is influenced by interactions with the external magnetic field as well as by random scatterings with thermal light partons. The presence of QCD matter comprising charged quarks can extend the lifetime and strength of the magnetic field, thereby enhancing the degree of heavy quark polarization. However, the random scatterings with QCD matter tend to diminish heavy quark polarization. In this study, we utilize the Landau-Lifshitz-Gilbert (LLG) equation to investigate both these contributions. Taking into account the realistic evolutions of medium temperatures and the in-medium magnetic fields at the Relativistic Heavy-Ion Collider (RHIC) and the Large Hadron Collider (LHC), we observe that heavy quark polarization is limited by the short lifetime of the magnetic field and the high temperatures of the medium. Furthermore, we explore the mass dependence of quark polarization, revealing that the polarization degree of strange quarks is much larger than that of charm quarks.

\end{abstract}
\date{\today}
 \maketitle
 
\section{Introduction}
In relativistic heavy-ion collisions, a deconfined state of matter comprising quarks and gluons is thought to be formed~\cite{Bazavov:2011nk}. Over the past decades, extensive studies have been conducted to explore the signals and properties of the Quark-Gluon Plasma (QGP)~\cite{Aoki:2006we,Rapp:2008tf,Braun-Munzinger:2015hba,Cao:2020wlm,Zhao:2020jqu,Song:2007ux,Shen:2014vra}, revealing it to be a nearly perfect fluid generated at the Relativistic Heavy-Ion Collider (RHIC)~\cite{Song:2010mg}. 
In 1986, the unusual suppression of charmonium $J/\psi$ was initially proposed as one of the indicators of QGP formation in nuclear collisions~\cite{Matsui:1986dk}. Unlike light hadrons, which are typically produced on the hypersurface close to the medium hadronization, heavy quarks are predominantly created in the initial parton hard scatterings, making them a relatively clean probe of the early-stage characteristics of the hot deconfined medium~\cite{Liu:2010ej}.

In non-central nuclear collisions, the substantial angular momentum carried by the two colliding nuclei can be transferred to the QGP medium~\cite{Liang:2004ph}. The rotation of the QGP not only leads to the directed flows of light and heavy flavor hadrons but also induces global spin polarization of particles in the rotating medium. In Au+Au collisions at $\sqrt{s_{NN}}=62.4$ and 200 GeV, the STAR Collaboration has experimentally determined the global polarization of $\Lambda$ hyperons~\cite{STAR:2017ckg}, showing a decrease with increasing collision energy compared to model predictions. The vortical effects on $\Lambda$ have been investigated using hydrodynamic models~\cite{Karpenko:2016jyx} and transport models~\cite{Li:2017slc}. The spin-orbital coupling can induce the polarization of quarks via particle scatterings~\cite{Zhang:2019xya,Weickgenannt:2020aaf,Liu:2020ymh,Gao:2020vbh}.
The directed flows of light and heavy flavor hadrons have been observed in experiments, which supports the significant effect of vortical fields in heavy ion collisions~\cite{STAR:2014clz, Bozek:2010bi,Becattini:2015ska,Chatterjee:2017ahy,STAR:2019clv}.

Moreover, the slight difference in the polarization of $\Lambda$ and $\bar \Lambda$ suggests the effect of the magnetic field, which can generate opposing polarizations for $\Lambda$ and $\bar \Lambda$ due to their opposite electric charges~\cite{Becattini:2016gvu}. 
When considering heavy quarks, their production can be influenced by the strong magnetic field in the elementary process~\cite{Chen:2024lmp}. Owing to the combined effects of the magnetic field and the stochastic spin-spin scatterings within the QGP, heavy quarks undergo precession around the magnetic field and experience spin polarization towards the field's direction~\cite{ref-LLG-1,ref-LLG-2}. Concurrently, the random scatterings lead to a stochastic distribution of quark spins, evolving towards a detailed balance. The degree of spin polarization for heavy quarks depends on the magnetic field's strength and the intensity of particle spin-spin random scatterings, which can be parameterized using the Landau–Lifshitz–Gilbert (LLG) equation~\cite{ref-LLG-3}.
The realistic evolutions of the QGP and magnetic fields in RHIC Au-Au collisions and LHC Pb-Pb collisions serve as the background for the LLG equation. 

This work is organized as follows: Section II provides an introduction to the LLG equation concerning heavy quark spin polarization, also discussing the in-medium magnetic field in the QGP. In Section III, the Langevin equation for heavy quark momentum evolution is introduced, elucidating the final momentum distribution of heavy quarks. Section IV gives the calculation of heavy quark spin polarization in both constant and realistic magnetic fields. Additionally, it examines the mass dependence of quark spin polarization in the fast-decaying magnetic field. Finally, Section V presents final conclusions.

\section{Landau–Lifshitz–Gilbert equation}
\begin{figure}[!htb]
\includegraphics[width=0.2\textwidth]{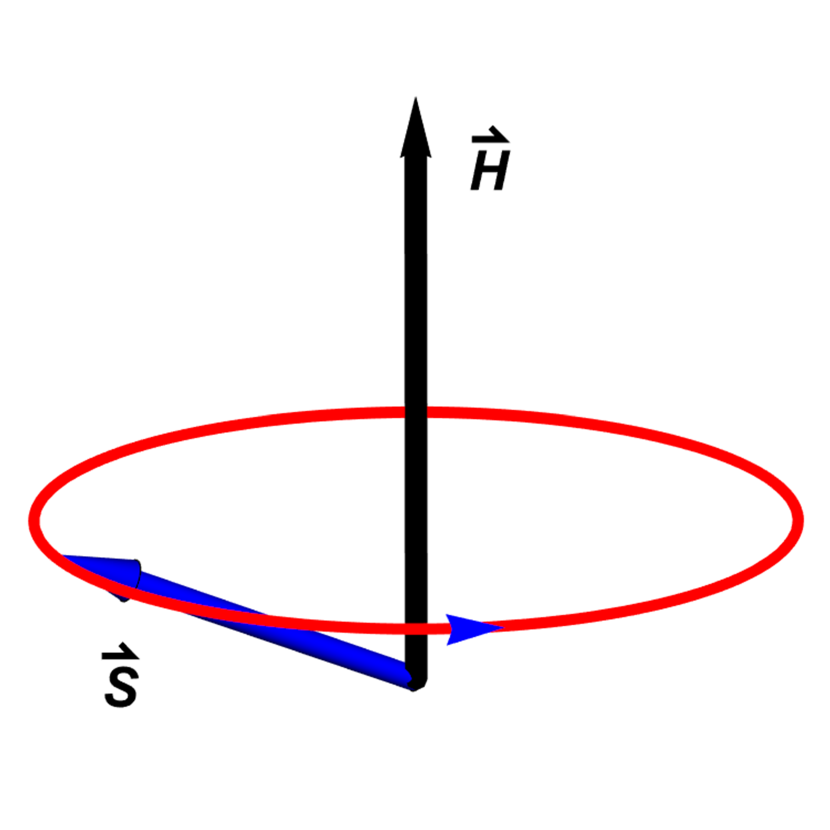}
\includegraphics[width=0.2\textwidth]{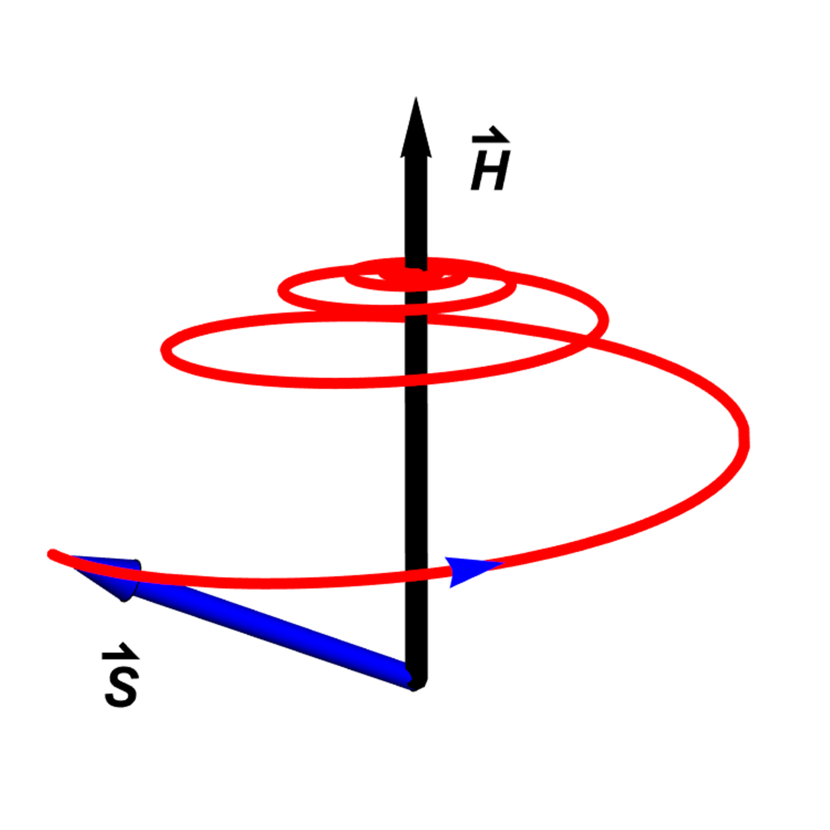}
\caption{Fermion spin evolution: (a) in the presence of a magnetic field and (b) when the polarization process is simulated considering the combined effects of spin-magnetic field interaction and spin-spin interactions.
}
\label{lab-fig1}
\end{figure}

Fermions move around the magnetic field direction as a result of the interplay between spin and the magnetic field. In the absence of additional interactions, no spin polarization occurs, as illustrated in Fig.\ref{lab-fig1}. However, upon factoring in spin-spin interactions among multiple fermions within the medium, the fermions' spins can align towards the magnetic field direction, influenced by lower Zeeman energies.

In heavy ion collisions, two nuclei approach each other at nearly the speed of light before colliding. In semi-central collisions, both the QGP comprising quarks and gluons and strong magnetic fields generated by spectator protons from the two nuclei can emerge. The presence of a strong magnetic field and spin-spin random scatterings can lead to the polarization of quark spins.
Given the rapid decay of the magnetic field during heavy ion collisions, it could imprint effects on heavy quarks produced at the initial stages of nucleus collisions. In the realm of condensed matter physics, the Landau-Lifshitz-Gilbert (LLG) equation has been extensively employed to explore the spin polarization of fermions within dense mediums under magnetic fields~\cite{ref-LLG-1,ref-LLG-2,ref-LLG-3},

\begin{align}
\frac{d {\bf S}}{dt} = &-\frac{\gamma}{1+\alpha ^2}{\bf S \times (H + H_{th})}  \nonumber \\
&-\frac{ \alpha\gamma}{1+\alpha ^2}{\bf S} \times [{\bf S \times (H + H_{th})}].
\label{eq-llg}
\end{align}
The vector ${\bf S}={\bf \mu}/|{\bf \mu}|$ represents a unit vector, where $\mu=g{\bf J_s}q/(2m_q)$ denotes the magnetic moment. Here, $|{\bf J}_s|=1/2$ signifies the spin vector of quarks, with $m_q$ and $q$ representing the mass and electric charge of the quark, respectively.
The gyromagnetic ratio $\gamma=\mu/J_s=gq/(2m_q)$ denotes the ratio of the magnetic moment to the spin of quarks, where the value of g-factor is $g = 2$~\cite{Ferrara:1992yc}. The vector ${\bf H}$ corresponds to the external magnetic field.
The damping factor, denoted by $\alpha$, characterizes the polarization process of quark spin and is influenced by both particle spin-spin random collisions and spin-magnetic field interactions. In subsequent computations, varying values of the damping factor will be considered to explore the maximal impact of the decaying magnetic field on heavy quark spin polarization in heavy ion collisions. Additionally, the impact of spin-spin random scatterings can be simulated by incorporating a randomly fluctuating magnetic field ${\bf H}_{th}$ into the LLG equation. The amplitude and time correlation of the fluctuating magnetic field are connected with the characteristics of the random scatterings.
The average value of ${\bf H}_{th}$ over time is zero, and its strength depends on the medium temperature. ${\bf H}_{th}$ is regarded as a white noise term that satisfies the following relation~\cite{ref-LLG-3,ref-LLG-4} in natural units,
\begin{align}
    \langle H_{th, i}(t)\rangle &= 0 \\
    \langle H_{th, i}(t) H_{th,j}(t^\prime)\rangle &={2\alpha T\over |{\bf \mu}| \gamma} \delta_{ij}\delta(t-t^\prime).
    \label{eq-thermal}
\end{align}
In the expression $H_{th,i}$, $i=(1,2,3)$ represents the index denoting the three components of ${\bf H}_{th}$ along the (x, y, z) directions. The inclusion of the fluctuating ${\bf H}_{th}(t)$ term in ${\bf S} \times ({\bf H} + {\bf H}_{th})$ results in the evolution of the quark's spin towards various directions at different time intervals. The term involving the damping factor $\alpha$ directs the quark spin to evolve towards the direction of the magnetic field, establishing a detailed balance between the processes of quark spin polarization and relaxation.
The strength of the fluctuating ${\bf H}_{th}$, which simulates the effects of spin-spin random interactions, increases with medium temperature.
The value of the damping factor can be studied using linear response theory. In this work, we will explore different values of $\alpha=(0.1, 1)$~\cite{ref:damping-factor-prl} to evaluate the maximum impact of the magnetic field on heavy quark spin polarization in nuclear collisions at RHIC and LHC. This effect approaches its maximum when $\alpha$ is around $1$. \tred{In relativistic heavy-ion collisions, the short lifetime of the realistic magnetic field ${\bf H}(t)$ will delay the spin polarization of quarks. Besides, when the mass of quarks becomes larger, the value of $\gamma$ in Eq.(\ref{eq-llg}) is smaller, which affects the degree of spin polarization for different quarks. }

\begin{figure}[!htb]
\includegraphics[width=0.3\textwidth]{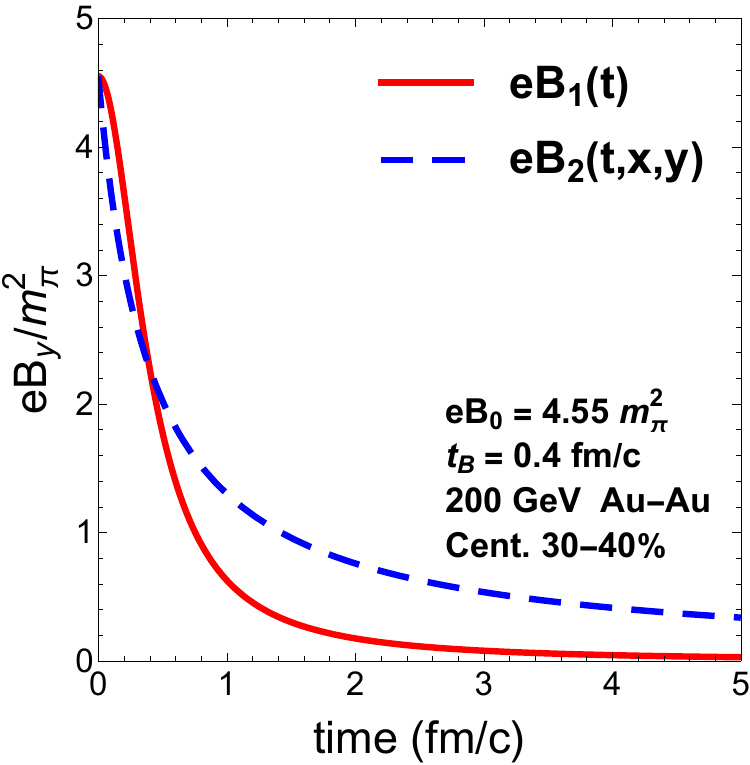}
\includegraphics[width=0.33\textwidth]{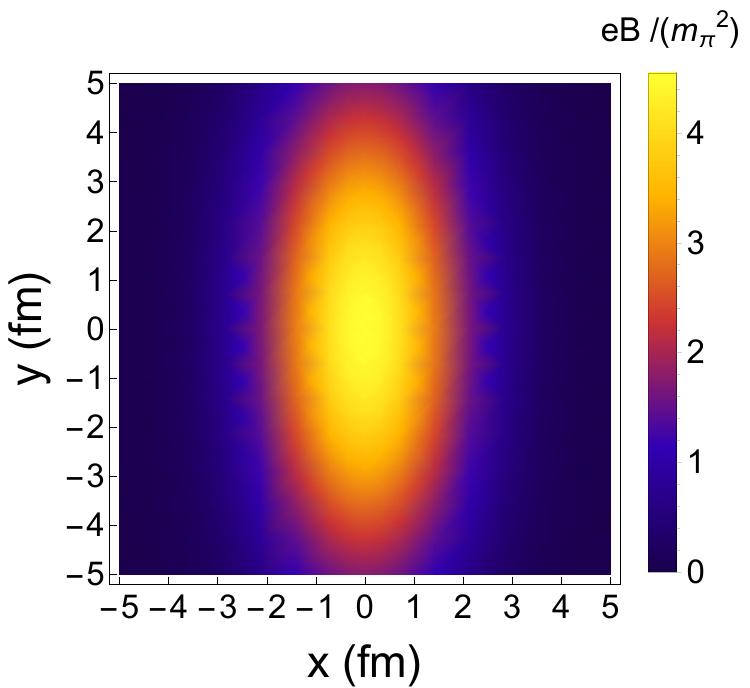}
\caption{ (Upper panel) Time evolution of the realistic in-medium magnetic field in RHIC 200 GeV Au-Au collisions in centrality 30-40\%. Solid and dashed lines are for $eB_1(t)$ and $eB_2(t,x,y)$ respectively with $(x=0, y=0)$. \tred{The value of $eB_0$ is determined based on the formula in~\cite{Deng:2012pc} by taking the corresponding value of the impact parameter in cent.30-40\%. $t_B$ is taken as 0.4 fm/c. (Lower panel) The spatial transverse distribution of the magnetic field $eB_2(t,x,y)$ at the initial time $t=0$. }
}
\label{lab-fig-Bfield}
\end{figure}

Another crucial factor influencing quark spin polarization is the time evolution of the external magnetic field ${\bf H}(t)$ produced during heavy ion collisions. In natural unit, the magnetic field ${\bf H}$ in Eq.(\ref{eq-llg}) is replaced with ${\bf B}$. 
Previous research has indicated that the QGP, composed of charged quarks, can become magnetized, and extend the lifetime of the magnetic field. This magnetic field within the medium is denoted as ${\bf H}(t)$ in the LLG equation. Different parameterizations of the in-medium magnetic fields have been studied~\cite{Jiang:2022uoe,Hongo:2013cqa,K:2022pzc,Huang:2017tsq}. We take two kinds of in-medium magnetic fields without and with spatial dependence, respectively from those references:
\begin{align}
    eB_1(t) &= {eB_0\over 1+ ({t\over t_B})^2} \\
    eB_2(t,x,y) &={eB_0\over 1+ ({t\over t_B})} \exp(-{x^2\over 2\sigma_x^2} - {y^2\over 2\sigma_y^2})
    \label{eq-Bmedium}
\end{align}
where $t_B$ represents the effects of medium on the magnetic field. $eB_0$ is the magnitude of the magnetic field generated by the moving spectator protons at time zero. \tred{It depends on the impact parameter and can be calculated with the reference~\cite{Deng:2012pc}.} The value is approximately $100m_\pi^2$ in LHC 5.02 TeV Pb-Pb collisions. This magnitude decreases almost to zero over a timescale of $2R_A/(\gamma_L c)$ as spectator protons leave the collision region. Here, $R_A$ represents the nuclear radius, and $\gamma_L$ is the Lorentz contraction factor. Different values of the $t_B$ indicate different magnetization of QGP~\cite{Yee:2013cya}, which is still under debate without determination. It is estimated to be $t_B=0.4$ fm/c~\cite{Huang:2017tsq}. \tred{A uniform distribution over positions is parametrized in $eB_1(t)$, while the spatial dependence are included in $eB_2(t,x,y)$. The parameters 
$\sigma_x=0.8(R-b/2)$ and $\sigma_y=0.8\sqrt{R^2-(b/2)^2}$ characterize the spatial distribution of the magnetic field, with $R=6.38$ fm~\cite{Hongo:2013cqa}. $b$ is the impact parameter varying with the collision centrality. 
The $eB_1$ decays faster than the $eB_2$. 
In the upper panel of Fig.\ref{lab-fig-Bfield}, 
the time evolution of in-medium magnetic fields are plotted, where the positions are taken as (x=0, y=0) for the magnetic field $eB_2(t,x,y)$. $eB_0=4.55\ m_\pi^2$ is calculated by taking a proper value of the impact parameter in the centrality 30-40\% in $\sqrt{s_{NN}}=200$ GeV Au-Au collisions~\cite{Deng:2012pc}. Two magnetic fields decrease significantly in the period $t<1.0$ fm/c. The lifetime of the magnetic field is much shorter than the lifetime of QCD matter. The spin polarization of heavy quarks by magnetic fields can be partially washed out by the random scatterings with thermal partons. Both magnetic fields will be considered in the evolution of charm and strange quarks at RHIC energy. }

\section{Langevin equation for momentum evolution}
When heavy quarks move inside the QGP, they dump energy to the medium via random scatterings and medium-induced radiation. As 
heavy quarks experience different degrees of spin polarization in different time scales, it is essential to provide a realistic description of the momentum evolution of heavy quarks. 
Assume a small momentum transfer in each random scattering, the motion of heavy quarks can be approximated to be Brownian motion. \tred{With the presence of the magnetic field, a Lorentz force also changes the momentum of heavy quarks~\cite{Das:2016cwd,Jiang:2022uoe}.} Accordingly, the non-relativistic form of the Langevin equation for the evolution of heavy quark momentum is expressed as follows~\cite{Cao:2013ita,He:2013zua,Chen:2017duy,Chen:2021akx}:
 \begin{align}
\label{lan-gluon}
{d{\bf p}\over dt}= -\eta(p) {\bf p} +{\bf \xi} + {\bf f}_g+{{\bf p}\over E_Q}\times (q {\bf B}).
\end{align}
On the right-hand side, random scatterings between heavy quarks and the thermal medium are approximated using the drag force alongside a noise term. \tred{The magnetic field contributes a Lorentz force acting on the heavy quarks with the electric charge $q$. ${\bf B}$ is the magnetic field given in previous sections.} The drag coefficient $\eta(p) = \kappa / (2TE_Q)$ depends on the medium temperature and the energy of the heavy quark $E_Q = \sqrt{p^2 + m_Q^2}$. For charm quarks, we consider the mass to be $m_c = 1.5$ GeV. The momentum diffusion coefficient $\kappa$ and the spatial diffusion coefficient $\mathcal{D}_s$ adhere to the relationship $\kappa \mathcal{D}_s = 2T^2$. Lattice QCD computations have determined the spatial diffusion coefficient at zero momentum and zero baryon chemical potential to be $\mathcal{D}_s 2\pi T \approx 2$ with $N_f = 2+1$~\cite{Altenkort:2023oms} in the temperature range $T_c < T < 2T_c$. However, values of $\mathcal{D}_s 2\pi T$ derived from phenomenological model assessments like T-matrix~\cite{Liu:2017qah}, Bayesian analysis~\cite{Xu:2017obm}, and the Langevin model~\cite{Cao:2015hia,Yang:2023rgb} are larger and exhibit variability within a broad range of $2.0 \lesssim \mathcal{D}_s 2\pi T \lesssim 7.0$, demonstrating a noticeable dependence on temperature~\cite{Dong:2019unq,Rapp:2018qla}. Its value has also been investigated using the deep learning approach with Convolutional Neural Network (CNN)~\cite{Guo:2023phd}, elucidating the experimental data of B hadrons wherein the dominant roles are played by the drag and noise terms rather than the radiation process. The parametrized spatial diffusion coefficient, exhibiting dependencies on both temperature and momentum, is extracted from the deep learning model as $D_s 2\pi T = 4.87 + 4.16(T/T_c -1)$ in QGP~\cite{Guo:2023phd}. 

The ${\bf \xi}$ term is approximated with a three dimensional white noise. When neglecting the momentum dependence, the time correlation of the white noise term satisfy,
\begin{align}
    \langle \xi^i(t) \xi^j(t^\prime)\rangle = \kappa \delta^{ij}\delta(t-t^\prime)
\end{align}
where $\xi^i$ (i=1,2,3) are the three components of the vector noise term. $t$ is the time. 
The radiation term is characterized as ${\bf f}_g = -d{\bf p}_g/dt$, with the momentum of the emitted gluon being randomly chosen based on the probability in the time interval $t \sim t + \Delta t$ as $P_{\rm rad}(t,\Delta t) = \Delta t \int dx d k_T^2 {dN_g / dx dk_T^2 dt}$~\cite{Guo:2000nz,Zhang:2003wk}. Here, $k_T$ represents the transverse momentum of the emitted gluon, while its spectrum is computed using perturbative QCD. The parameter $x = E_g / E_Q$ signifies the ratio of gluon energy to heavy quark energy. The initial momentum distribution of heavy quarks can be produced via  fixed-order
plus next-to-leading log formula (FONLL) calculation~\cite{Cacciari:1998it,Cacciari:2001td}. The initial spatial density of heavy quarks is proportional to the density of nuclear binary collisions.

The QGP formed in heavy ion collisions has been established as a strongly coupled medium, effectively described by the hydrodynamic equations~\cite{Schenke:2010rr}. The MUSIC package has been utilized to describe the time and spatial evolutions of the hot medium~\cite{Schenke:2010nt}, with initial temperature profiles determined based on the final charged hadron multiplicity. The equation of state (EoS) for the deconfined medium is parameterized using the lattice EoS at zero baryon density and the EoS of the hadron resonance gas~\cite{Bernhard:2016tnd,HotQCD:2014kol}. We monitor the evolution of heavy quarks until the local temperature of the medium decreases to a decoupling temperature of $T=170$ MeV, following which the hadronization of heavy quarks takes place through either the coalescence process~\cite{Greco:2003vf,Greco:2003mm} or the fragmentation process~\cite{Back:2002wb,Fries:2003vb,Qin:2015srf}.
Considering the polarization of heavy quarks, the spin-coalescence model should be utilized to examine the spin polarization of the final hadrons~\cite{Yang:2017sdk,Sheng:2020ghv}. This study focuses on the maximal impact of the magnetic field on heavy quark spin polarization, which is found to be minor due to the rapid decay of the magnetic field and significant spin-spin random scatterings. Therefore, we do not provide the final distributions of heavy flavor hadrons; instead, we present the polarization of heavy quarks before their hadronization process.
\tred{Magnetic fields not only directly influence the momentum and spin of heavy quarks~\cite{Das:2016cwd}, as demonstrated in this study, but also impact the evolution of light quarks and the bulk medium, such as the chiral magnetic effect~\cite{Kharzeev:2015znc,Kharzeev:2010gd,Shi:2017cpu,Huang:2015oca}. Magneto-hydrodynamics can be employed to study the effect of the magnetic field in the evolution of the bulk medium~\cite{Inghirami:2016iru}. As heavy quarks encounter the strongest magnetic field in the early stage of heavy ion collisions, we have temporarily neglected this impact on hydrodynamics and the equation of state of the medium when investigating heavy quark dynamics.}

\section{Spin polarization in magnetic field}
\subsection{Polarization in constant magnetic field}
\begin{figure}[!htb]
\includegraphics[width=0.23\textwidth]{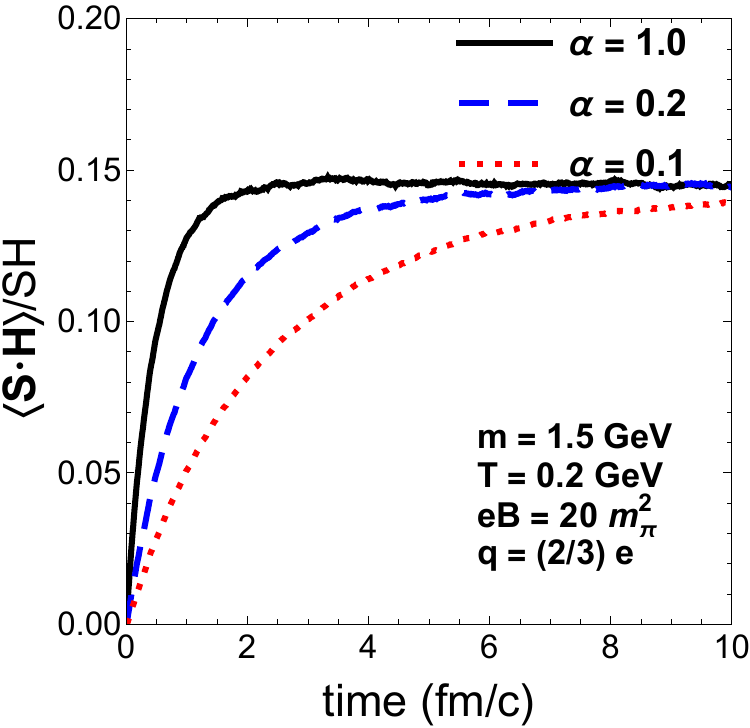}
\includegraphics[width=0.23\textwidth]{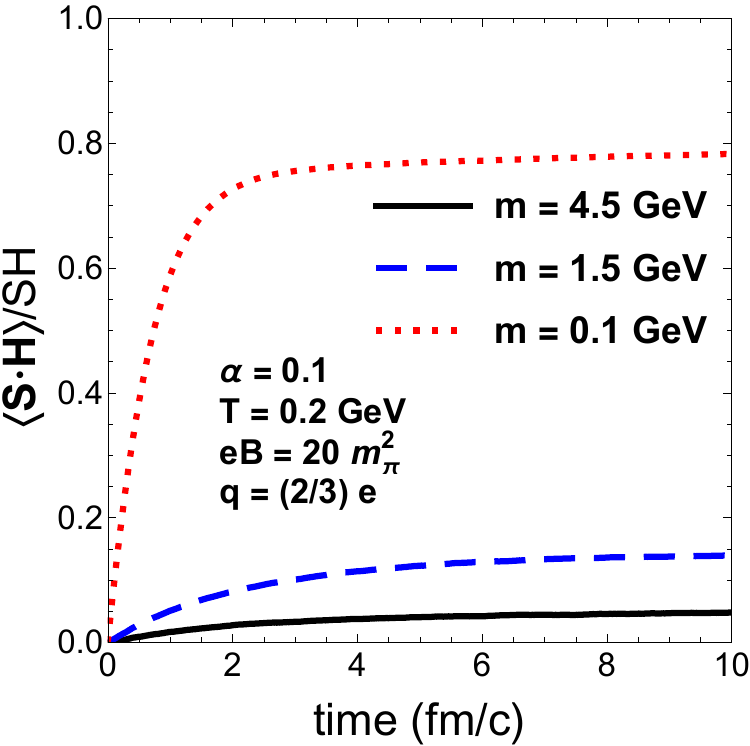}
\caption{ Left subplot: The spin polarization of heavy quarks over time is examined under the conditions of a constant temperature $T=0.2$ GeV and a constant magnetic field strength of $eB=20 m_\pi^2$. Various values of the damping factor $\alpha=(0.1, 0.2, 1.0)$ are utilized to investigate the rate of quark spin polarization.
Right subplot: Different masses m=(0.1, 1.5, 4.5) GeV which characterize the cases of strange, charm, and bottom quarks, are employed to analyze the extent of quark spin polarization. The damping factor in the LLG equation is taken as $\alpha=0.1$. The electric charge of the quark is taken as $q=(2/3)e$.
}
\label{lab-alpha-mass}
\end{figure}

Firstly, we study the spin polarization of heavy quarks in the hot static medium with constant temperature and constant magnetic field. In Fig.\ref{lab-alpha-mass}, we take the medium temperature to be $T=0.2$ GeV, and the electric charge of charm quarks is $q=(2/3)e$. To calculate the degree of spin polarization along the magnetic field, we introduce the normalized variable, $\langle {\bf S}\cdot {\bf H}\rangle/(SH)$. ${\bf S}$ and ${\bf H}$ are the spin and magnetic fields respectively. $\langle...\rangle$ represents the average over a large number of quarks. Its value is between 0 and 1, which corresponds to the cases of random distribution and complete spin polarization. In Fig.\ref{lab-alpha-mass} various damping factors $\alpha$ are utilized to accommodate distinct rates of spin polarization, although they do not alter the ultimate equilibrium of spin polarization, as depicted by the three lines in the left subplot of Fig.\ref{lab-alpha-mass}. The behavior reveals that with differing $\alpha$ values, the level of spin polarization during the time interval $0<t<2$ fm/c varies, closely aligning with the duration of the magnetic field's presence in heavy ion collisions. Therefore, 
diverse damping factors can yield varying final spin polarizations of quarks during heavy ion collisions. 
When the medium temperature becomes higher, the impact of random scatterings leads to a reduction in the degree of spin polarization.

Even heavy quarks are produced at the beginning of nuclear collisions where the magnetic field is the strongest, the large mass of charm quarks may delay their spin polarization. We employ different values of quark mass by taking bottom, charm, and strange quarks in the LLG equation. The degree of spin polarization becomes smaller when quark mass becomes larger, please see the right subplot of Fig.\ref{lab-alpha-mass}. Note that if without the thermal fluctuating magnetic field ${\bf H}_{th}$, all the lines will approach the unit when time is large enough. 

\subsection{Polarization in heavy-ion collisions}
\begin{figure}[!htb]
\includegraphics[width=0.35\textwidth]{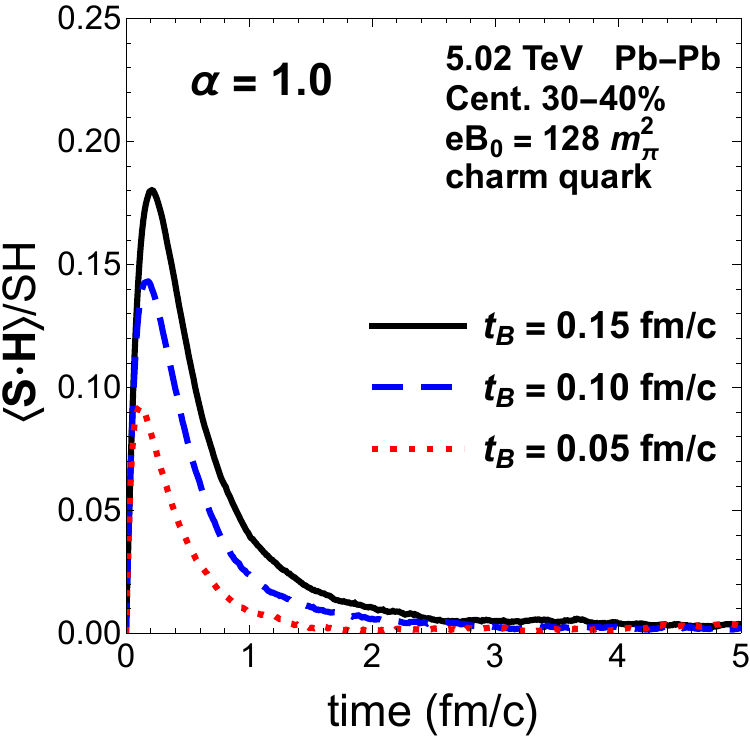}
\caption{The average polarization of quark spins over time for centrality 30-40\% in the central rapidity of 5.02 TeV Pb-Pb collisions. Three distinct lines represent varying levels of in-medium magnetic fields with the form $eB_1(t)$ by taking different values of $t_B$. The initial maximum magnetic field strength is set at $eB_0=128m_\pi^2$ at $t=0$.
}
\label{lab-HIC-time}
\end{figure}

After studying the speed and degree of quark spin polarization in the static medium, we now shift our focus to the time evolution of heavy quarks within a realistic magnetic field and hot medium generated during nuclear collisions. In 5.02 TeV Pb-Pb collisions within centrality 30-40\%, we evolve the charm quark's spin and momentum with the LLG equation and Langevin equation respectively. A uniformly distributed magnetic field $eB_1(t)$ is employed. The average polarization of charm quark spins with the magnetic field is plotted at each time step in Fig.\ref{lab-HIC-time}. The magnetic field is strongest in the early stage. Charm quarks are polarized at first, and then they undergo subsequent random collisions. The spin polarization in the early stage, will be washed out by the thermal random interactions in QGP. Various values of $t_B$ are utilized to explore distinct levels of in-medium magnetic fields in Fig. \ref{lab-HIC-time}. Even charm quarks experience different degrees of spin polarization in the early stage of the magnetic field, this polarization information will be washed out by the hot medium. 

\begin{figure}[!htb]
\includegraphics[width=0.23\textwidth]{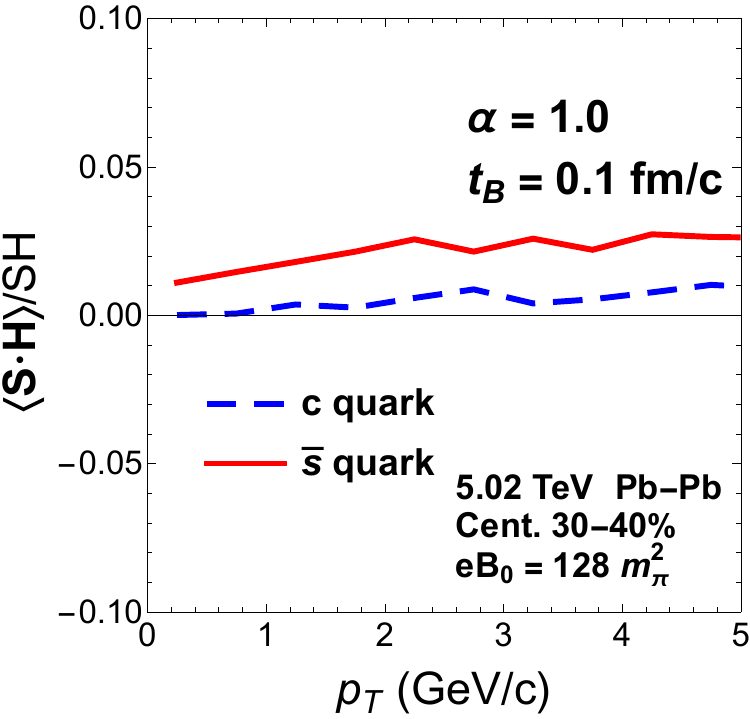}
\includegraphics[width=0.23\textwidth]{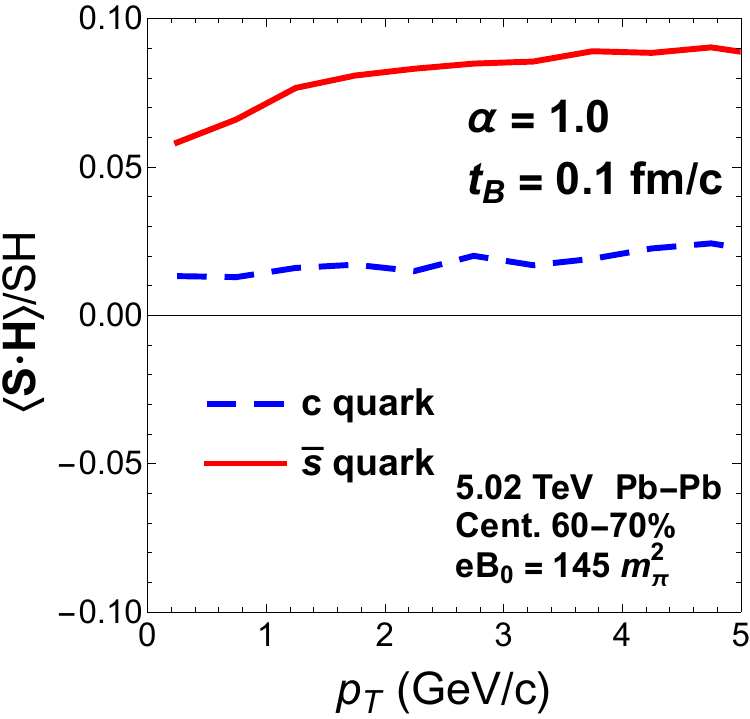}
\caption{ 
The final spin polarization of charm and strange quarks plotted against transverse momentum in 5.02 TeV Pb-Pb collisions in the centrality 30-40\% (left subplot) and 60-70\% (right subplot).  The damping factor is taken as $\alpha=1.0$. A uniformly distributed magnetic field $eB_1(t)$ is taken, where the value of $eB_0$ is determined to be $128\ m_\pi^2$ and $145\ m_\pi^2$ respectively in two centralities. 
}
\label{lab-HIC-diffcent}
\end{figure}

The final polarization of charm and (anti-)strange quarks are plotted as a function of transverse momentum in Fig.\ref{lab-HIC-diffcent}. When charm quarks carry different transverse momentum, the time exposed in the magnetic field and hot medium will be different, which may alter the degree of their spin polarization. 
The contribution of the magnetic field and hot medium is also different in different collision centralities. Therefore, the final spin polarization of charm and strange quarks are plotted in Fig.\ref{lab-HIC-diffcent}. In cent.30-40\%, charm quarks are nearly unpolarized, while the polarization of strange quarks is only a few percent. Note that strange quark annihilation and creation from thermal partons have been neglected, as the Langevin and LLG equations do not consider particle production processes. In more peripheral collisions, like cent.60-70\%, the effect of the magnetic field is strong while the hot medium effect is much weaker. The degree of spin polarization is around $\sim 7\%$, please see the right panel of Fig.\ref{lab-HIC-diffcent}.

\begin{figure}[!htb]
\includegraphics[width=0.23\textwidth]{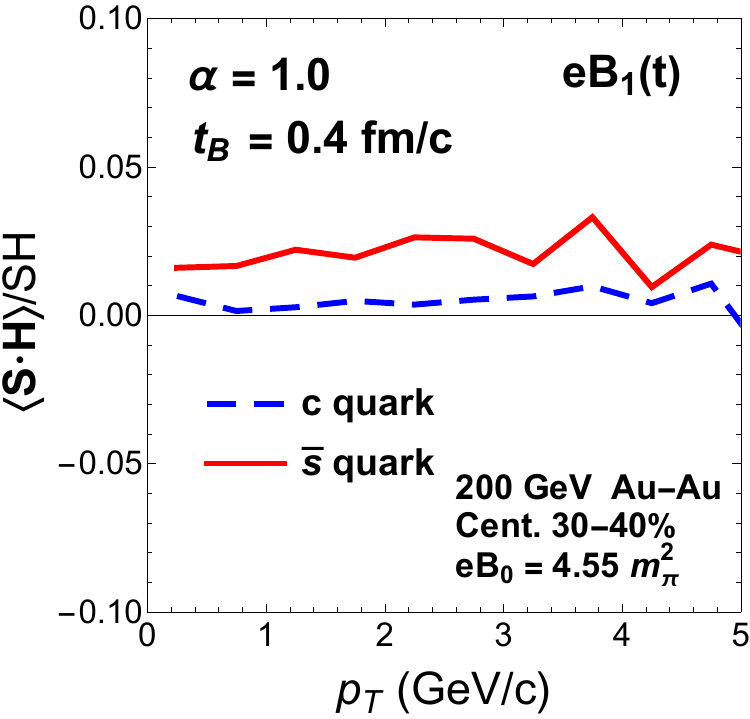}
\includegraphics[width=0.23\textwidth]{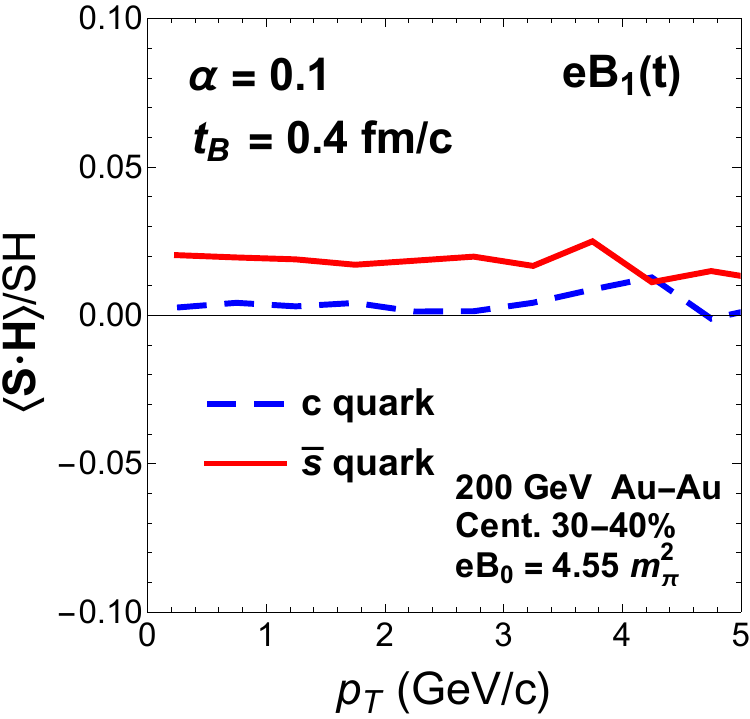}
\includegraphics[width=0.23\textwidth]{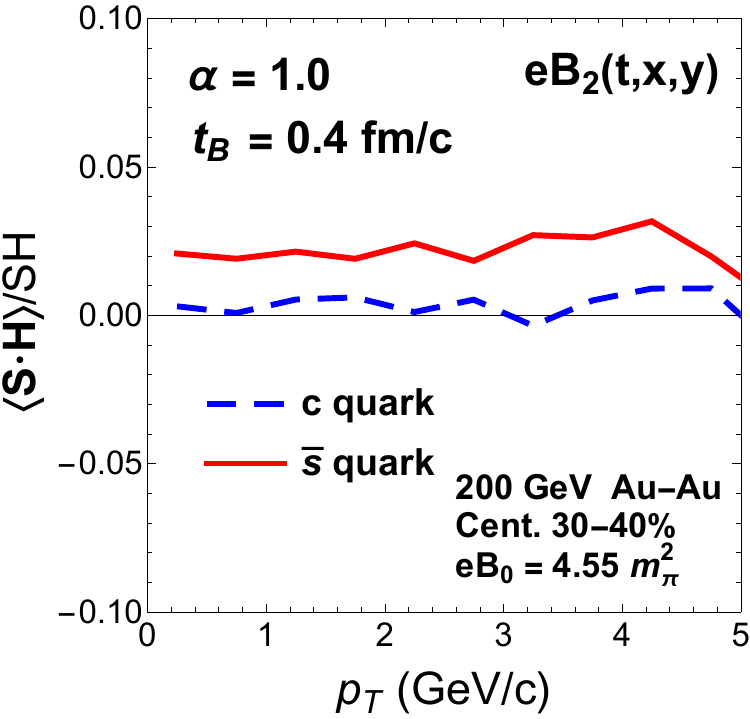}
\includegraphics[width=0.23\textwidth]{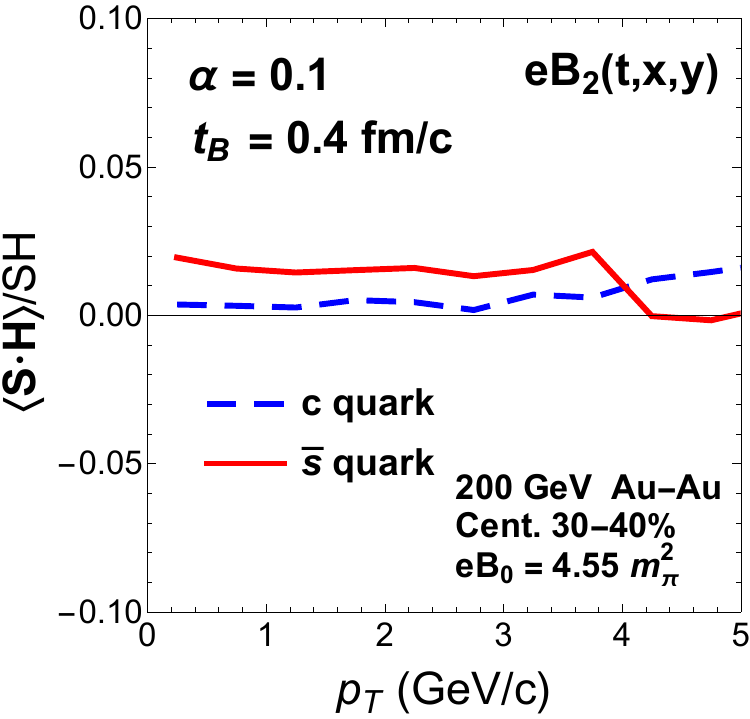}
\caption{ (upper panel) The average polarization of charm and anti-strange quark spins is plotted as a function of transverse momentum. A uniformly distributed magnetic field $eB_1(t)$ is employed. This analysis is conducted within centrality 30-40\% of RHIC 200 GeV Au-Au collisions. The magnetic field time scale is set at $t_B=0.4$ fm/c, with an initial maximum magnetic field strength determined to be $eB_0=4.55 m_\pi^2$ at $t=0$. Two different damping factor values, $\alpha=(1.0, 0.1)$, are utilized in the left and right subplots, respectively. (Lower panel) Similar to the upper panel, but incorporating the spatially dependent magnetic field $eB_2(t,x,y)$. 
}
\label{lab-RHIC-spin}
\end{figure}

At LHC energies, the degree of quark spin polarization is reduced by the high temperatures of the hot medium. Therefore, we also investigate the spin dynamics of charm and (anti-)strange quarks in the centrality 30-40\% of RHIC 200 GeV Au-Au collisions.
In Fig.\ref{lab-RHIC-spin}, the final average polarization of charm quark spins before hadronization is plotted. Two kinds of magnetic fields are employed in the upper and lower panels of Fig.\ref{lab-RHIC-spin} respectively. As shown in the figure, the degree of charm quark spin polarization is very weak without clear evidence of polarization induced by the magnetic field. The damping factor is set as $\alpha =1 $ and $0.1$ respectively in the left and right subplots, which do not change the result evidently. When employing a magnetic field with spatial dependence $eB_2(t,x,y)$, the final results do not change evidently, because of the short lifetimes of two magnetic fields. The spin polarization would be a bit stronger in more peripheral collisions where medium temperatures are lower. 

\tred{
The LLG equation incorporates spin polarization of heavy quarks induced by the external in-medium magnetic field and thermal random scatterings simultaneously. Due to the large mass of heavy quarks, their trajectories can be traced, while thermal production and annihilation of heavy quark pairs can be neglected. Therefore, both the LLG and Langevin equations can be employed to simulate the evolution of heavy quark spin and momentum. This approach offers a novel method to investigate the spin polarization of heavy flavor hadrons such as $J/\psi$, which has been observed in recent experiments.
The final production of $J/\psi$ in nucleus-nucleus collisions is dominated by the regeneration process, where most of $J/\psi$ are produced with partially spin-polarized charm quarks. Our phenomenological model provides both non-equilibrated spin and momentum distributions of charm quarks, essential for the realistic description of $J/\psi$ spin polarization in heavy-ion collisions. 
Furthermore, the polarization rate in the LLG equation, akin to the diffusion coefficient in the Langevin equation, offers an intuitive understanding of the interactions between heavy quark spin and the medium.
}

\section{Summary}
We investigate the spin polarization of heavy quarks within a hot medium and under the influence of a magnetic field. The interplay between spin-magnetic field interactions and spin-spin random scatterings drives quark spins toward the direction of the magnetic field. Simultaneously, random scattering induces the evolution of particle spins towards a randomized distribution. This phenomenon can be represented equivalently as a fluctuating magnetic field in the Landau–Lifshitz–Gilbert (LLG) equation. To model the evolution of quark spin, we employ the LLG equation, while the Langevin equation is utilized to describe quark momentum dynamics. Realistic magnetic field and temperature profiles of the medium are taken into account in 5.02 TeV Pb-Pb and 200 GeV Au-Au collisions. Even one can observe heavy quark polarization in a constant magnetic field within a hot medium, the magnitude of charm quark polarization is almost zero in heavy ion collisions where the magnetic field rapidly decays. In more peripheral collisions, the degree of spin polarization can be a bit stronger with only a few percent.
This implies that the spin polarization of charmonium or other heavy-flavor hadrons may be dominated by other mechanisms such as the vortical field.

\vspace{1cm}
\noindent {\bf Acknowledgement}: This work is supported by the National Natural Science Foundation of China
(NSFC) under Grant Nos. 12175165.

\end{document}